\documentclass[aps,prl,reprint,groupedaddress,amsmath,amssymb,showpacs]{revtex4-1}
\usepackage{graphicx}
\usepackage{CJK}
\pdfoutput=1
\begin{document}
\begin{CJK*}{UTF8}{}
\title{Coarse-Grained Modeling of a Deformable Nematic Vesicle}
\author{Jun Geng (\CJKfamily{gbsn}耿君)}
\author{Jonathan V. Selinger}
\author{Robin L. B. Selinger}
\affiliation{Liquid Crystal Institute, Kent State University, Kent, Ohio 44242, USA}
\date{December 18, 2011}
\begin{abstract}
We develop a coarse-grained particle-based model to simulate membranes with nematic liquid-crystal order. The coarse-grained particles form vesicles which, at low temperature, have orientational order in the local tangent plane. As the strength of coupling between the nematic director and the vesicle curvature increases, the vesicles show a morphology transition from spherical to prolate and finally to a tube. We also observe the shape and defect arrangement around the tips of the prolate vesicle.
\end{abstract}
\maketitle
\end{CJK*}

Over the past twenty-five years, a major theme of research in condensed-matter physics has been the complex interaction of geometry with orientational order and topological defects. Both theoretical and experimental studies have investigated order and defects on surfaces of \emph{fixed shape}, such as colloidal particles or droplets~\cite{Vitelli2006, Fernandez-Nieves2007, Shin2008, Skacej2008, Bates2008, Kralj2010, Lopez-Leon2011}. These studies have shown, for example, that a nematic phase on a spherical surface will form four defects of topological charge +1/2 each, and these defects may be exploited to develop colloidal particles that will self-assemble into tetrahedral lattices for photonic applications~\cite{Nelson2002}. Inspired by this potential application, many authors studied how to control the arrangement of the four half-charged defects in a nematic phase. Many effects have been considered, including elastic anisotropy~\cite{Shin2008}, external field~\cite{Skacej2008}, and curvature of the colloidal particles~\cite{Kralj2010}.

Further research has investigated orientational order and defects on \emph{deformable vesicles}, which serve as simple analogues for biological membranes~\cite{Park1992, Lubensky1992, Jiang2007, Ramakrishnan2010}. These studies show that defects in the orientational order will deform fluid vesicles into non-spherical shapes. For example, some vesicles have tilt order, which can be modeled as XY order in the local tangent plane; these vesicles will exhibit two defects of topological charge +1 each, and can deform into prolate or oblate shapes~\cite{Jiang2007}. Other vesicles composed of T-shaped lipids or surfactants with rod-like heads may have nematic order in the local tangent plane~\cite{Oda1999}. Theoretical studies have predicted that the four half-charged defects will induce these vesicles to deform into tetrahedra~\cite{Park1992}.

So far, theoretical studies of orientational order in deformable vesicles have considered systems that are idealized in several ways: at zero temperature, with no elastic anisotropy, with only certain couplings between orientational order and curvature, and with shapes that are slight perturbations on perfect spheres. A key question is whether vesicle shape would be qualitatively different if any of these simplifying assumptions were relaxed. Computer simulation provides a useful approach to this issue. For example, simulations can investigate problems where the geometry is not a perfect sphere but rather a more complex disordered shape, with bumps of positive and negative Gaussian curvature. One simulation method uses a triangulated-surface model with tangent-plane orientational order; this method has indeed shown complicated tube and inward tubulate shapes~\cite{Ramakrishnan2010}. However, a disadvantage of the triangulated-surface model is that the connectivity of the surface is fixed, unlike experimental systems in which molecules can detach from and rejoin the vesicles.

In this article, we develop an alternative method to simulate orientational order in deformable nematic vesicles, using a coarse-grained particle-based model. This method allows the particles to be assembled into a membrane with orientational order in the local tangent plane. The membrane spontaneously selects its own shape, which may be flat, spherical, or more complex. Furthermore, the interaction of the coarse-grained particles can be correlated with molecular features. Using this model, we calculate the arrangement of topological defects and the shape of the vesicle as a function of the interaction parameters. In particular, we find a morphology transition from spherical to prolate and finally to a tube as the coupling between nematic order and curvature increases.

To develop an appropriate simulation approach, we generalize a coarse-grained membrane simulation model \emph{without} tangent-plane order due to Ju Li and collaborators~\cite{Lykotrafitis, Liu2009, Zheng2010, Yuan2010}; see also Ref.~\cite{Ballone2006}. In their approach, a bilayer membrane is represented by a single layer of interacting coarse-grained point particles, each of which carries a polar vector degree of freedom $\mathbf{\hat{n}}$ representing the preferred layer normal direction. The interaction potential favors association of particles with $\mathbf{\hat{n}}$ vectors lying parallel and side-by-side. In simulations, the particles self-assemble into pancake-shaped single-layer aggregates showing liquid-like self-diffusion and membrane elasticity. When another term is added to the potential favoring a slight splay between the $\mathbf{\hat{n}}$ vectors of neighboring particles, particles spontaneously coalesce into spherical shells. Each particle represents not a single molecule but a large patch of membrane containing many molecules, and the surrounding solvent is implicit. This approach is more coarse-grained compared to other implicit-solvent lipid models in the literature~\cite{Bennun2009} as it does not separately resolve hydrophilic head/hydrophobic tail components across the width of a lipid bilayer.

\begin{figure}
\includegraphics[width=2.5in]{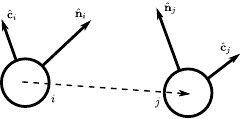}
\caption{\label{fig:twoVectorsModel} Each coarse-grained particle has a vector $\mathbf{\hat{n}}$, which aligns along the local membrane normal, and a vector $\mathbf{\hat{c}}$, which has nematic alignment within the local tangent plane.}
\end{figure}

To simulate a membrane \emph{with} tangent-plane order, for instance formed by T-shaped lipids, we generalize the preceding model by defining a coarse-grained point particle with \emph{two} vector degrees of freedom, as shown in Fig.~\ref{fig:twoVectorsModel}. The vector $\mathbf{\hat{n}}$ again defines the preferred layer normal direction, and a new vector $\mathbf{\hat{c}}$ represents the local nematic director orientation in the membrane's tangent plane. Both $\mathbf{\hat{n}}$ and $\mathbf{\hat{c}}$ are unit vectors and are always perpendicular to each other. Particles interact with each other via an anisotropic Lennard-Jones-type pairwise potential with a distance cut-off:
\begin{eqnarray}
\label{eq:uij}
&&u_{ij}(\mathbf{\hat{n}}_{i},\mathbf{\hat{n}}_{j},\mathbf{\hat{c}}_{i},\mathbf{\hat{c}}_{j},\mathbf{x}_{ij})=\\
&&\quad u_{R}(x_{ij})
+[1+\alpha[a(\mathbf{\hat{n}}_{i},\mathbf{\hat{n}}_{j},\mathbf{\hat{c}}_{i},\mathbf{\hat{c}}_{j},\mathbf{\hat{x}}_{ij})-1]]u_{A}(x_{ij}).\nonumber
\end{eqnarray}
In this expression, the repulsive and attractive parts of the potential are given by
\begin{eqnarray}
    \label{eq:ar}
&&u_{R}(r)=\varepsilon[(R_{cut}-r)/(R_{cut}-r_{min})]^8 \text{ for $r<R_{cut}$,}\nonumber\\
&&u_{A}(r)=-2\varepsilon[(R_{cut}-r)/(R_{cut}-r_{min})]^4 \text{ for $r<R_{cut}$,}\nonumber\\
&&u_{R}(r)=u_{A}(r)=0 \text{ for $r\ge R_{cut}.$}
\end{eqnarray}
Note that the repulsive and attractive terms have exponents 8 and 4, respectively, in contrast with the exponents 12 and 6 for the Lennard-Jones potential. These reduced exponents soften the potential and enhance the fluidity of the membrane. The coefficient $\alpha$ controls the strength of the anisotropic orientational interactions, which are defined by the function:
\begin{eqnarray}
\label{eq:alpha}
&&a(\mathbf{\hat{n}}_{i},\mathbf{\hat{n}}_{j},\mathbf{\hat{c}}_{i},\mathbf{\hat{c}}_{j},\mathbf{\hat{x}}_{ij})=\\
&&\quad 1-[1-(\mathbf{\hat{n}}_{i}\cdot\mathbf{\hat{n}}_{j})^2-\beta ]^2
-[(\mathbf{\hat{n}}_i \cdot\mathbf{\hat{x}}_{ij})^2-\gamma]^2 \nonumber\\
&&\quad -[(\mathbf{\hat{n}}_j \cdot\mathbf{\hat{x}}_{ij})^2-\gamma]^2
+2\eta^2 [(\mathbf{\hat{c}}_i \cdot \mathbf{\hat{c}}_j)^2 -1].\nonumber
\end{eqnarray}
In this function, $\beta=\sin^2(\theta_0)$ and $\gamma=\sin^2(\theta_0/2)$, where $\theta_0$ represents the preferred angle between the $\mathbf{\hat{n}}$ vectors of two neighboring particles.  As shown by Li and coworkers  \cite{Lykotrafitis}, if $\theta_0 = 0$,  particles tend to self-assemble into flat membranes with the $\mathbf{\hat{n}}$ vectors along the membrane normal. If $\theta_0 \neq 0$, then the $\mathbf{\hat{n}}$ vectors of neighboring particles align with a separation angle $\theta_0$; i.e. they favor a splay, and particles self-assemble into vesicles with spontaneous curvature.

The novel aspect of our potential is the $\eta^2$ term, which favors parallel or antiparallel alignment of the $\mathbf{\hat{c}}$ vectors of neighboring particles, giving rise to \emph{nematic liquid-crystal order} in the plane of the membrane. By varying the parameter $\eta$ at fixed temperature, we can vary the strength of the nematic order, and hence the Frank elastic constants of the liquid crystal and the strength of the coupling between curvature and orientational order.

We perform a series of simulations with about 10,000 particles over a range of $\eta$ from 0.2 to 0.5, at  temperature $k_BT \approx 0.22\varepsilon$.  Other parameters are
$d = 1.0$, $\epsilon = 1.0$, $\alpha = 3.1$, $r_{min} = 2^{1/6} d$, $R_{cut} = 2.55d$, and  $\theta_0 = 0.0375$ radians.
Initially, coarse-grained particles are distributed to cover the surface of a sphere of radius $30d$ using the random sequential absorption method~\cite{Widom1966}. The value of $\theta_0$ was selected to set the model membrane's spontaneous curvature to match the vesicle's initial radius. In the initial state, $\mathbf{\hat{n}}$ vectors are aligned radially and $\mathbf{\hat{c}}$ vectors are randomly oriented in the local tangent plane. The temperature is  increased and maintained by a Langevin thermostat.  The $\mathbf{\hat{c}}$ nematic director field quickly orders to form a population of $+/-$ half-charge defects; this texture coarsens via defect pair-annihilation until four positive half-charge defects remain. The vesicle then relaxes toward its equilibrium shape.

\begin{figure}
\begin{tabular}{cccc}
  $\eta$ & Front & Side & Top \\
  0.2 &
\includegraphics[width=0.96in]{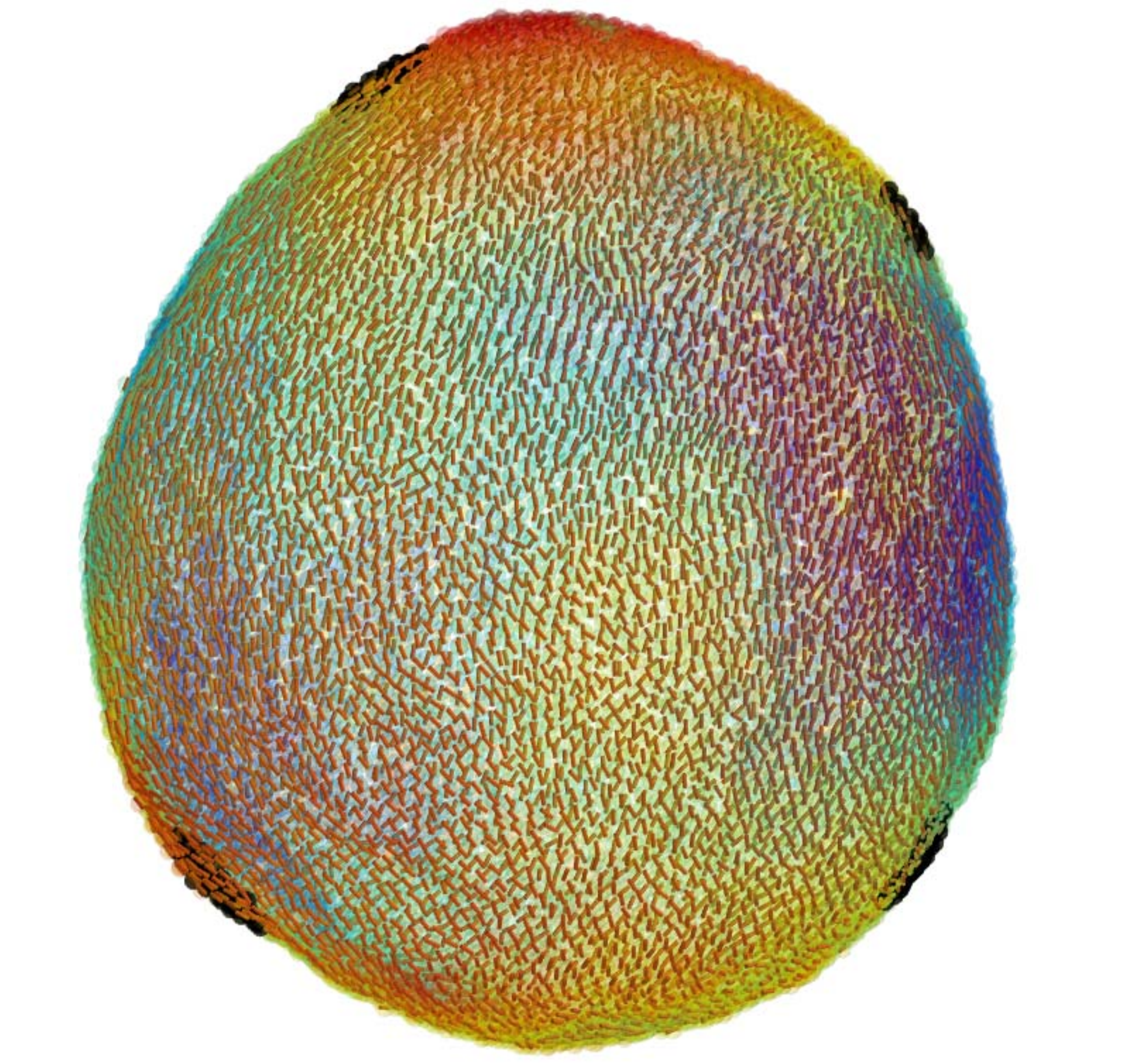} &
\includegraphics[width=0.96in]{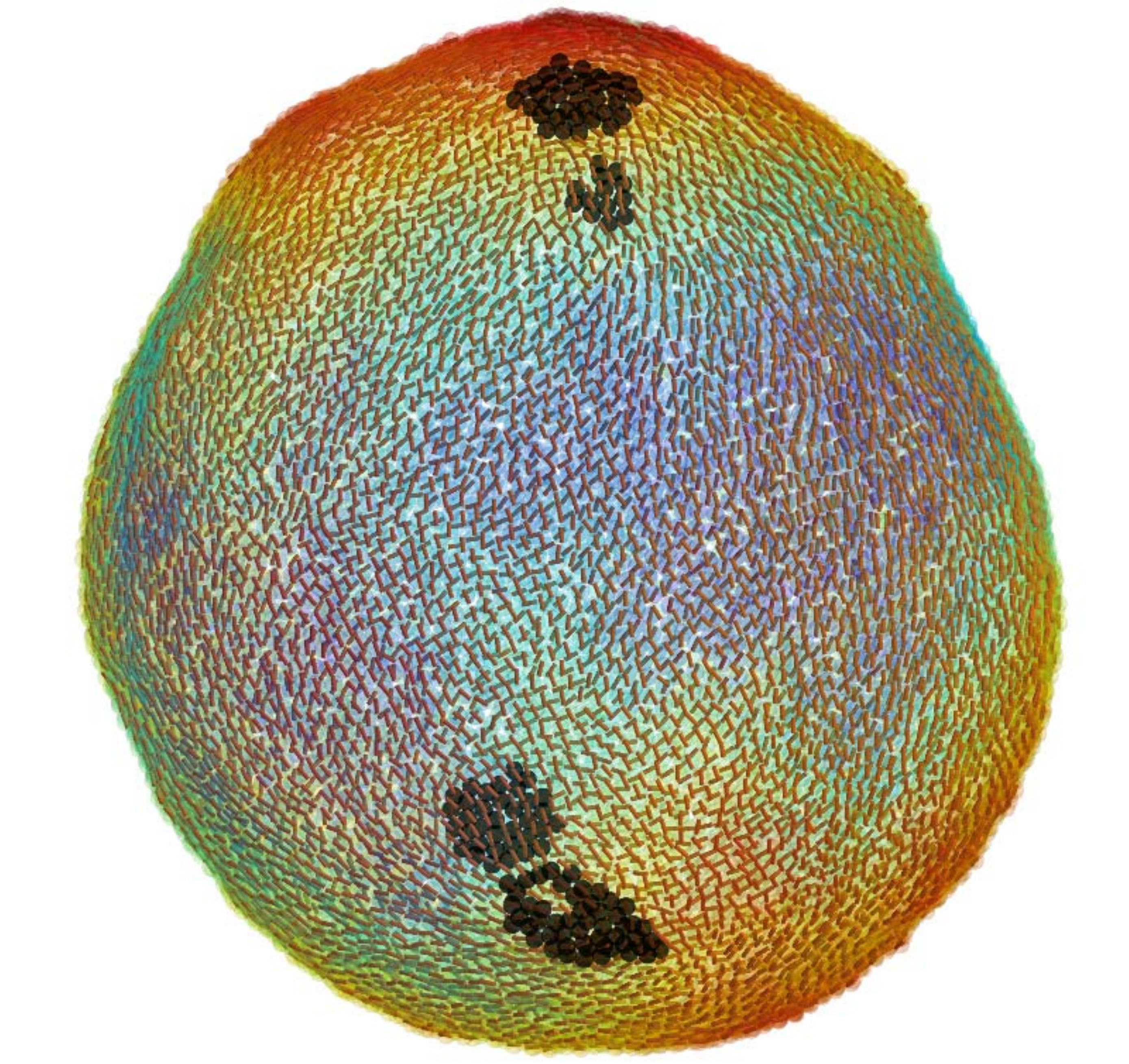} &
\includegraphics[width=0.96in]{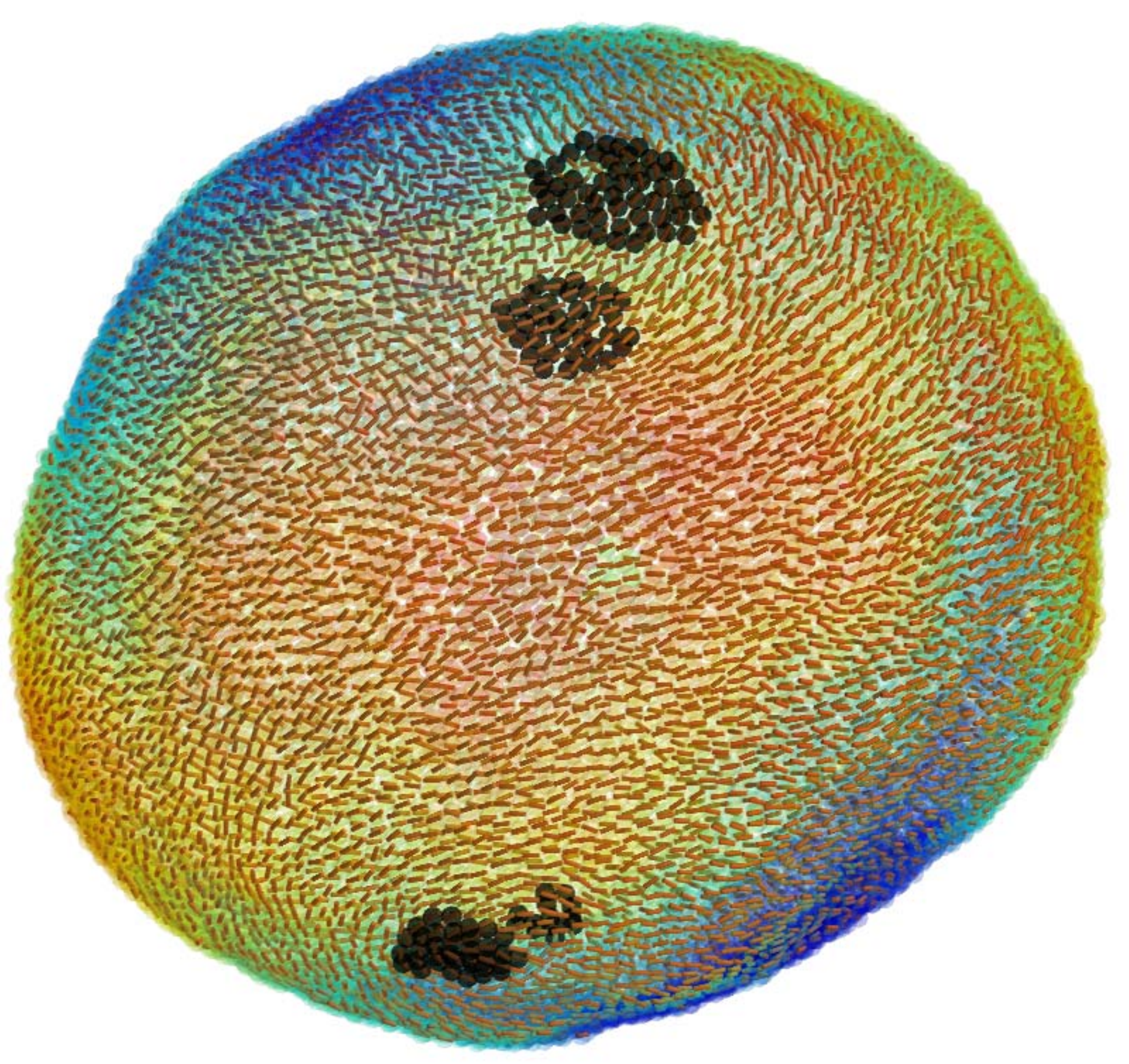} \\
  0.25 &
\includegraphics[width=0.96in]{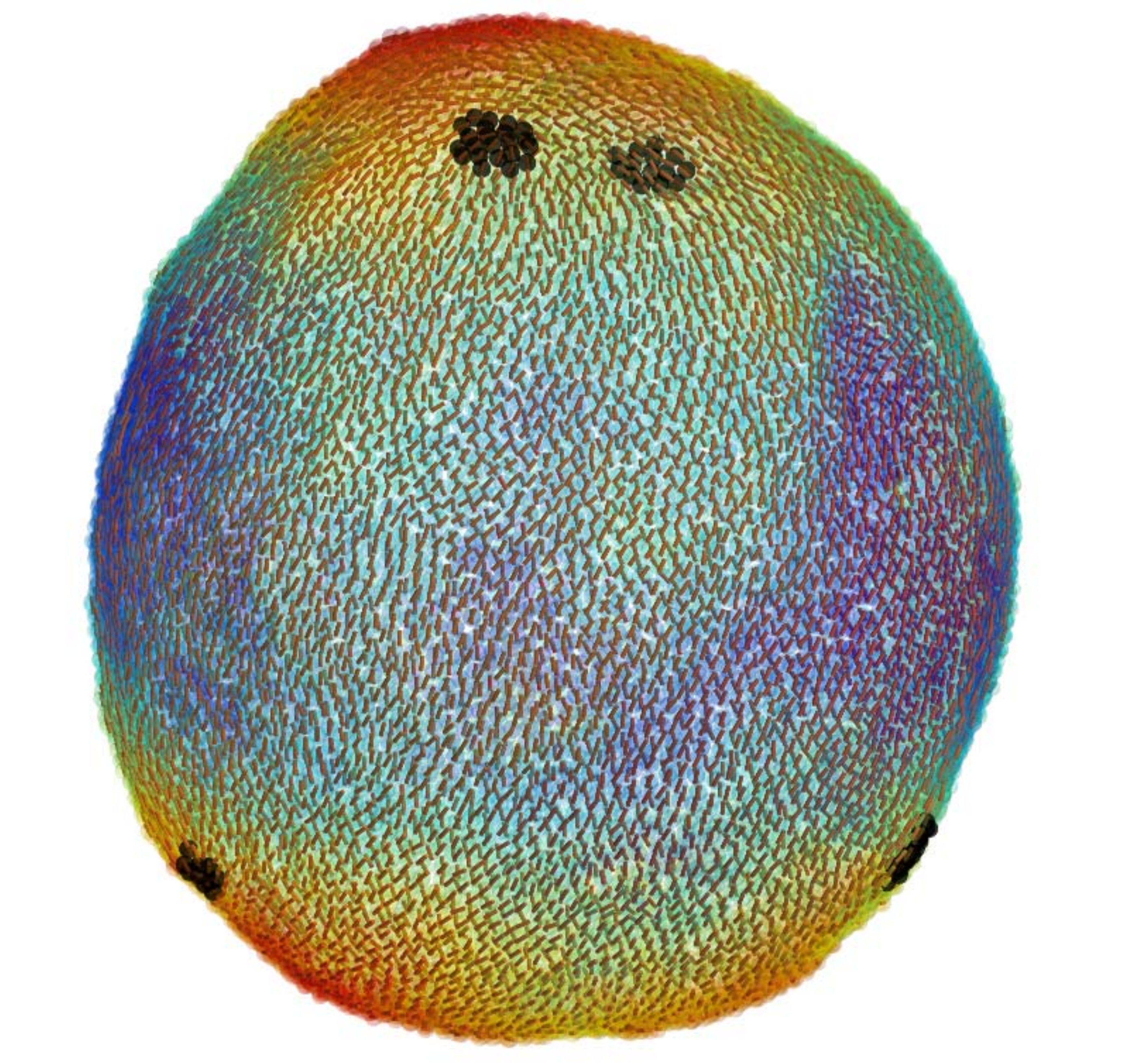} &
\includegraphics[width=0.96in]{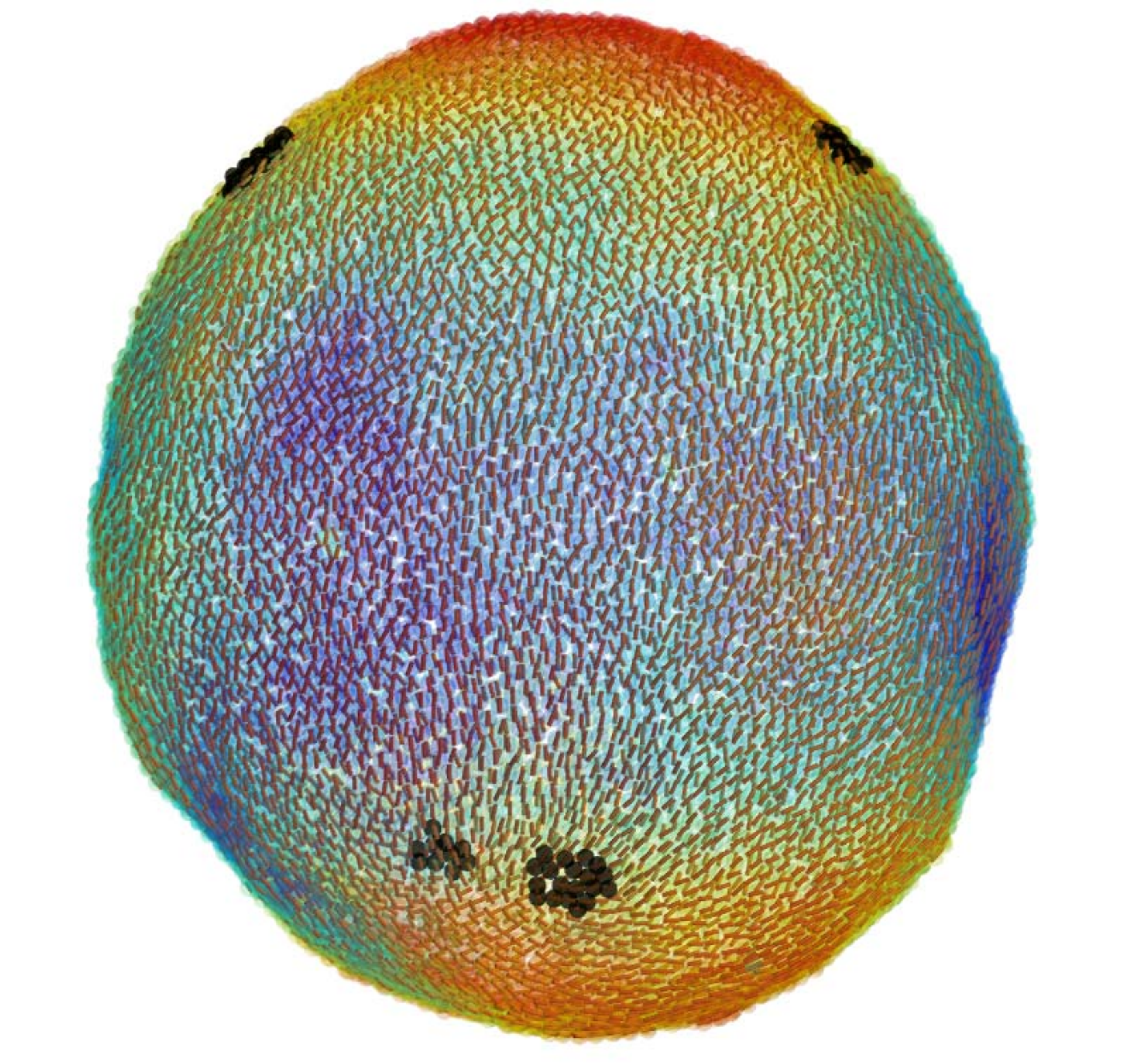} &
\includegraphics[width=0.96in]{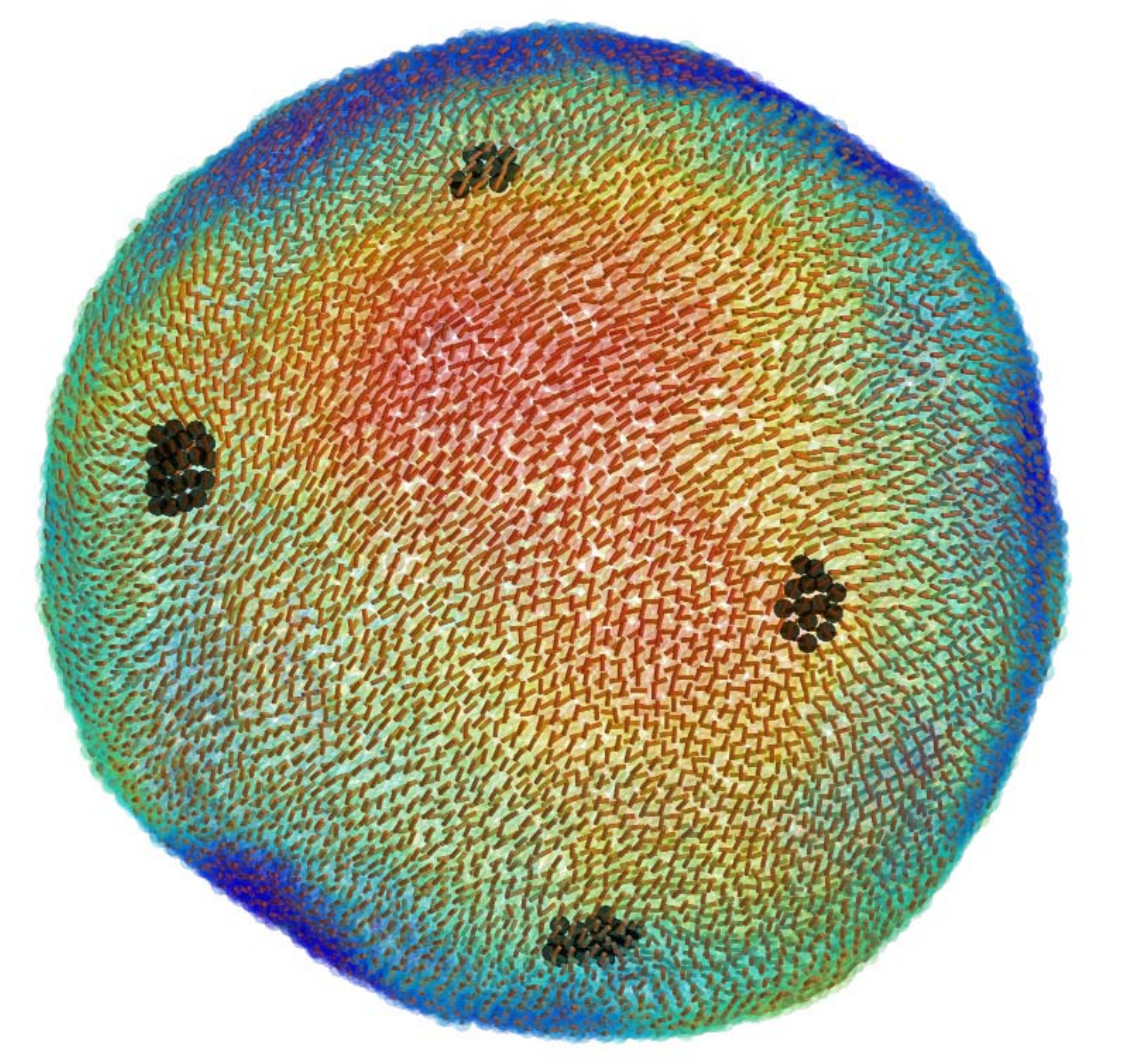} \\
  0.3 &
\includegraphics[width=0.96in]{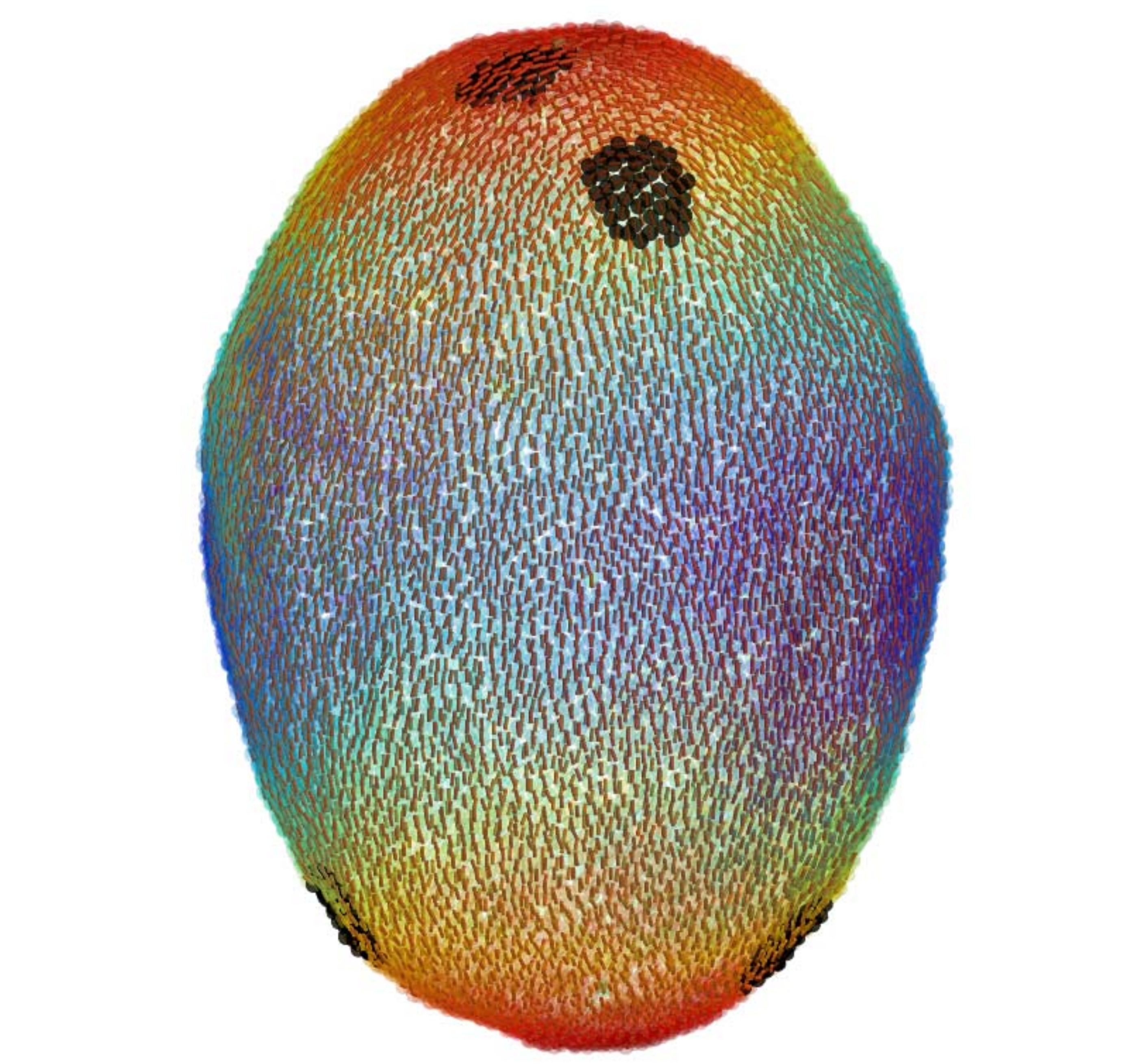} &
\includegraphics[width=0.96in]{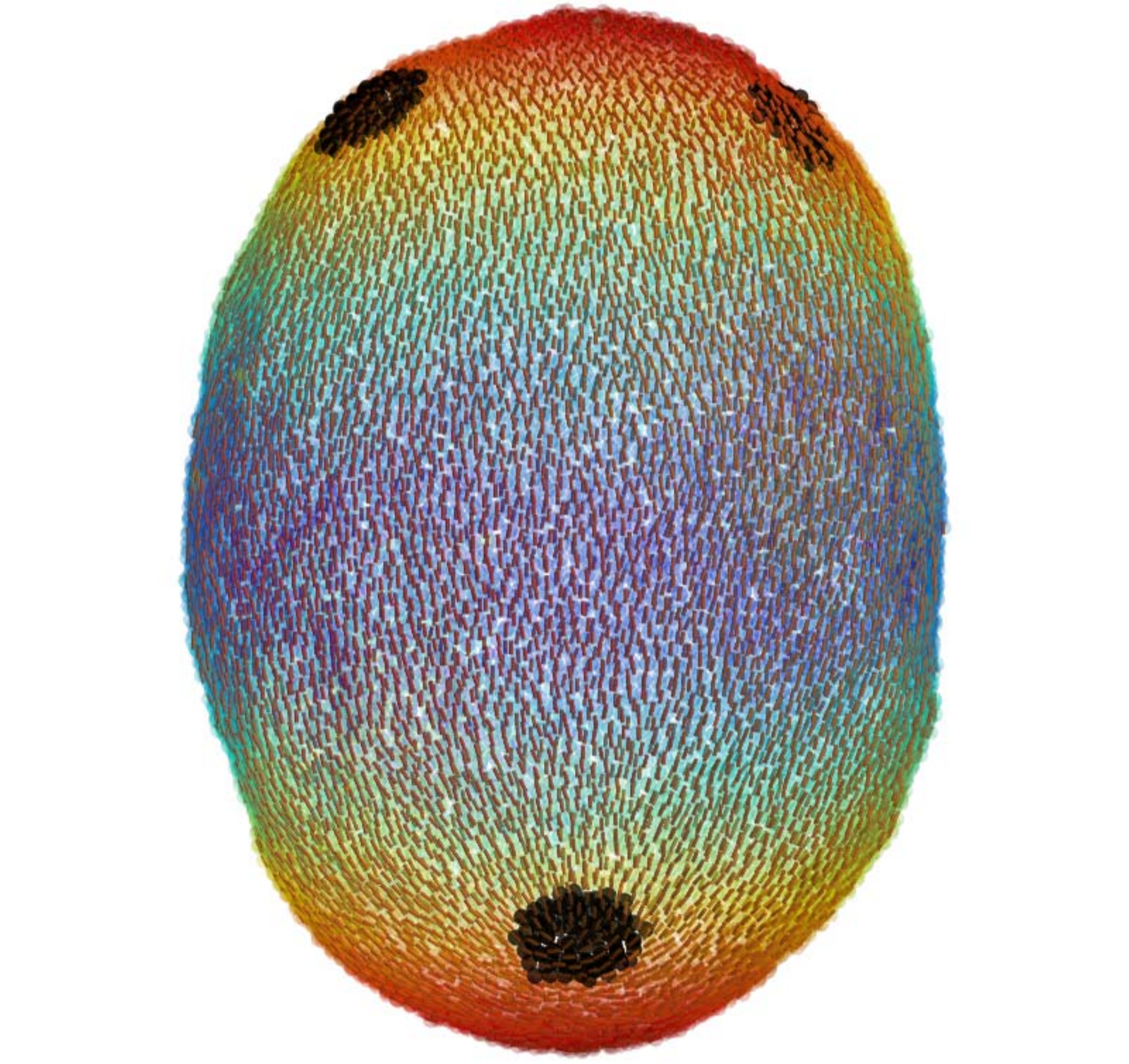} &
\includegraphics[width=0.96in]{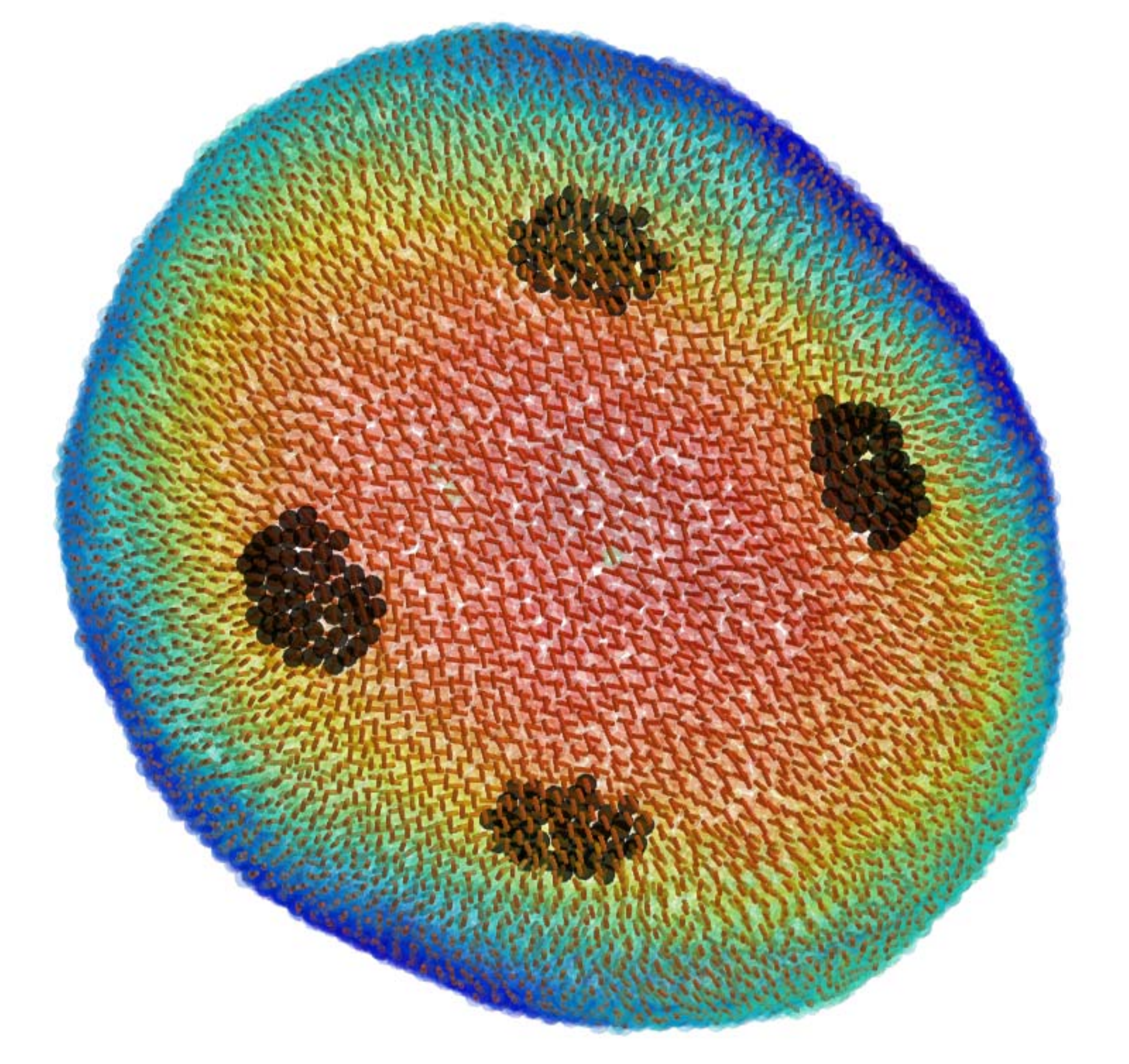} \\
  0.35 &
\includegraphics[width=0.96in]{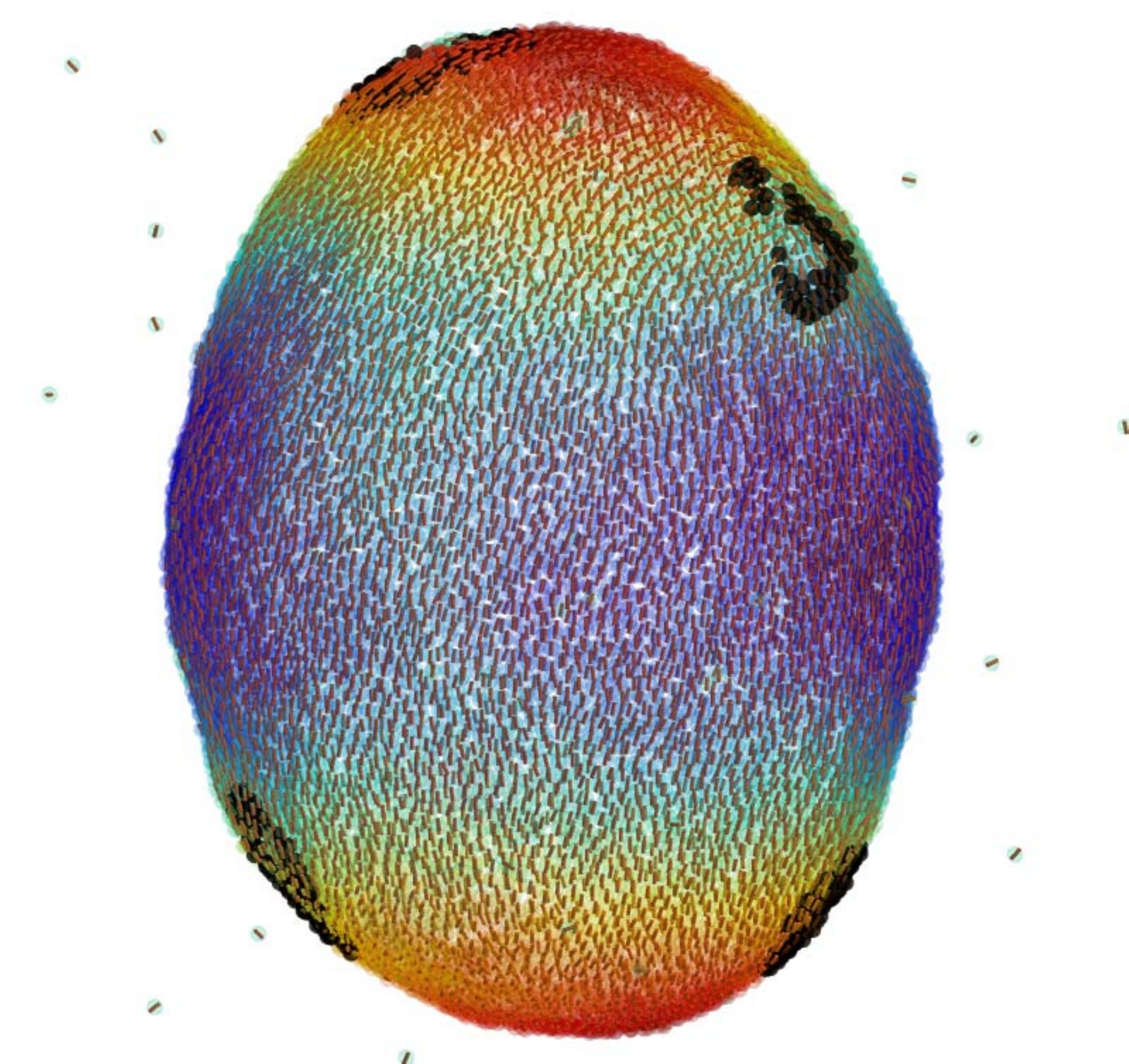} &
\includegraphics[width=0.96in]{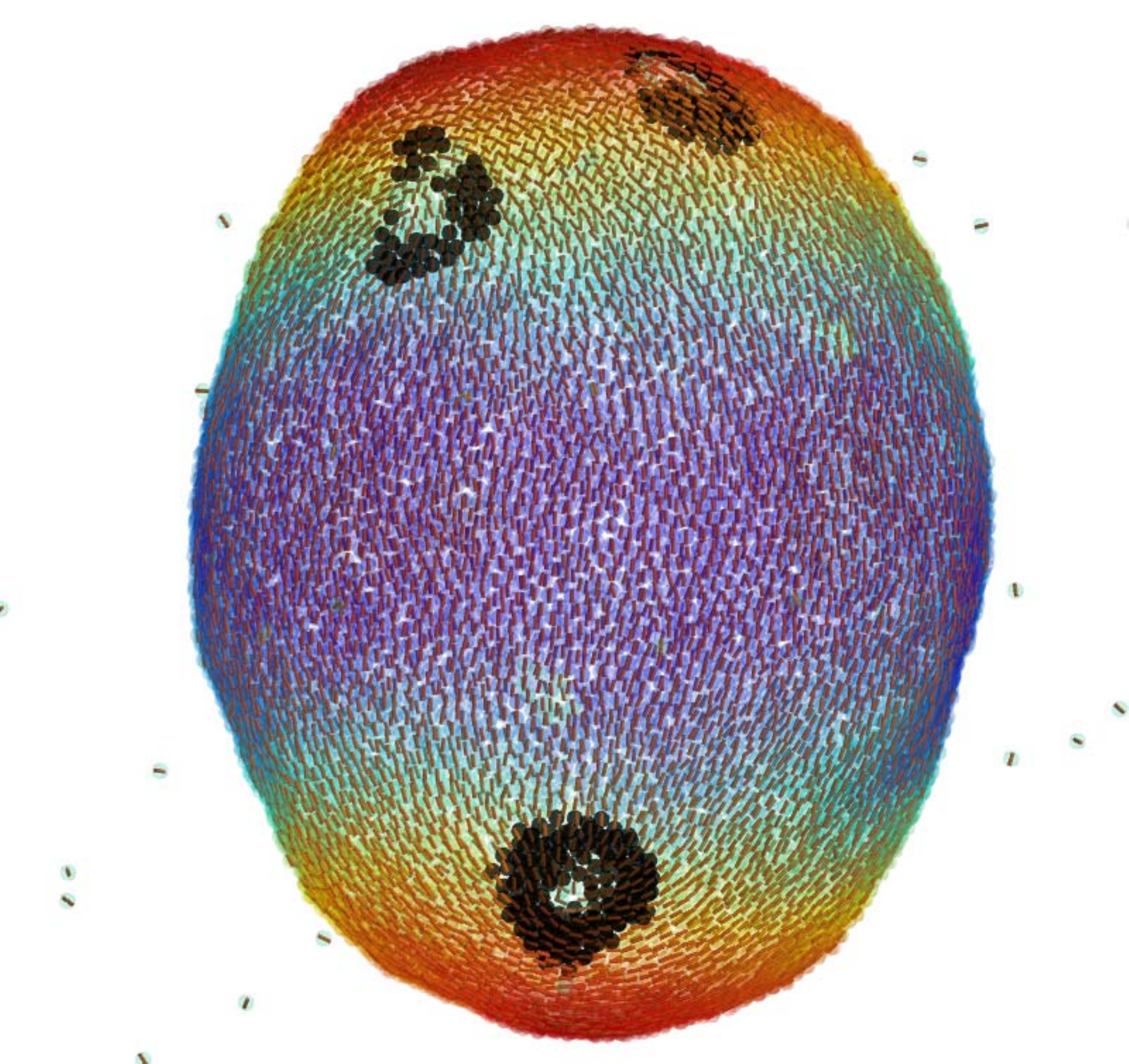} &
\includegraphics[width=0.96in]{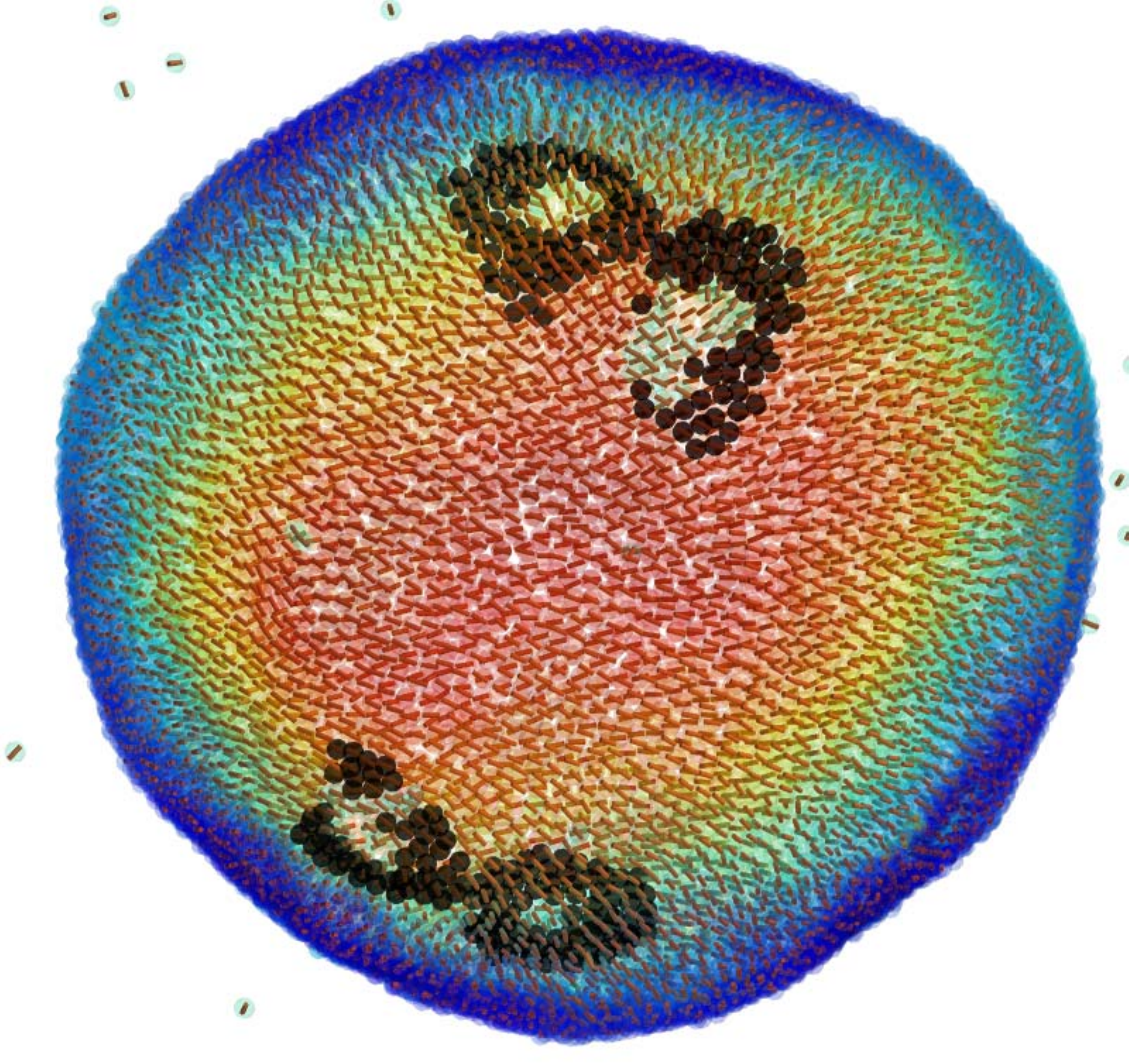} \\
  0.4 &
\includegraphics[width=0.96in]{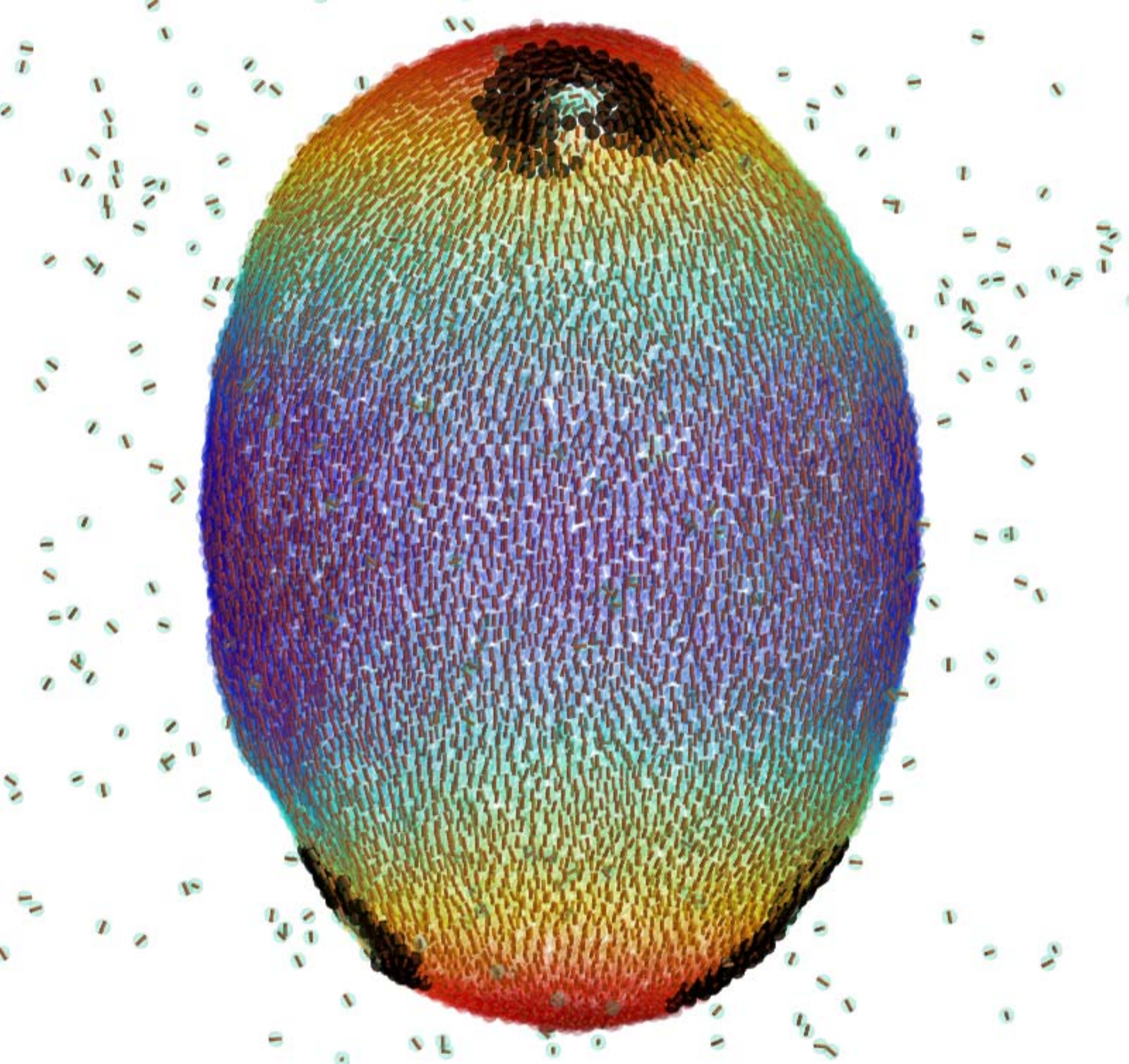} &
\includegraphics[width=0.96in]{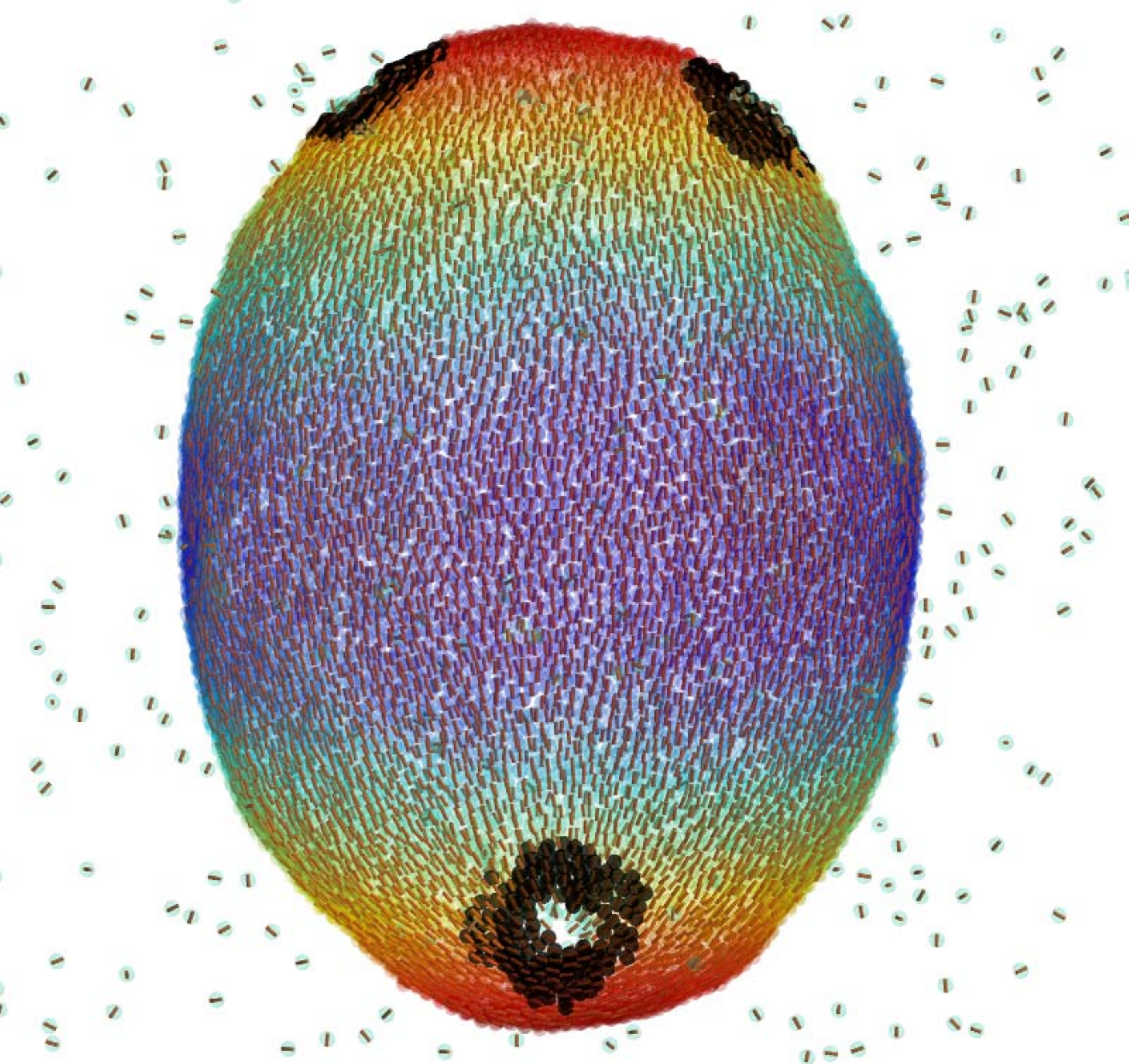} &
\includegraphics[width=0.96in]{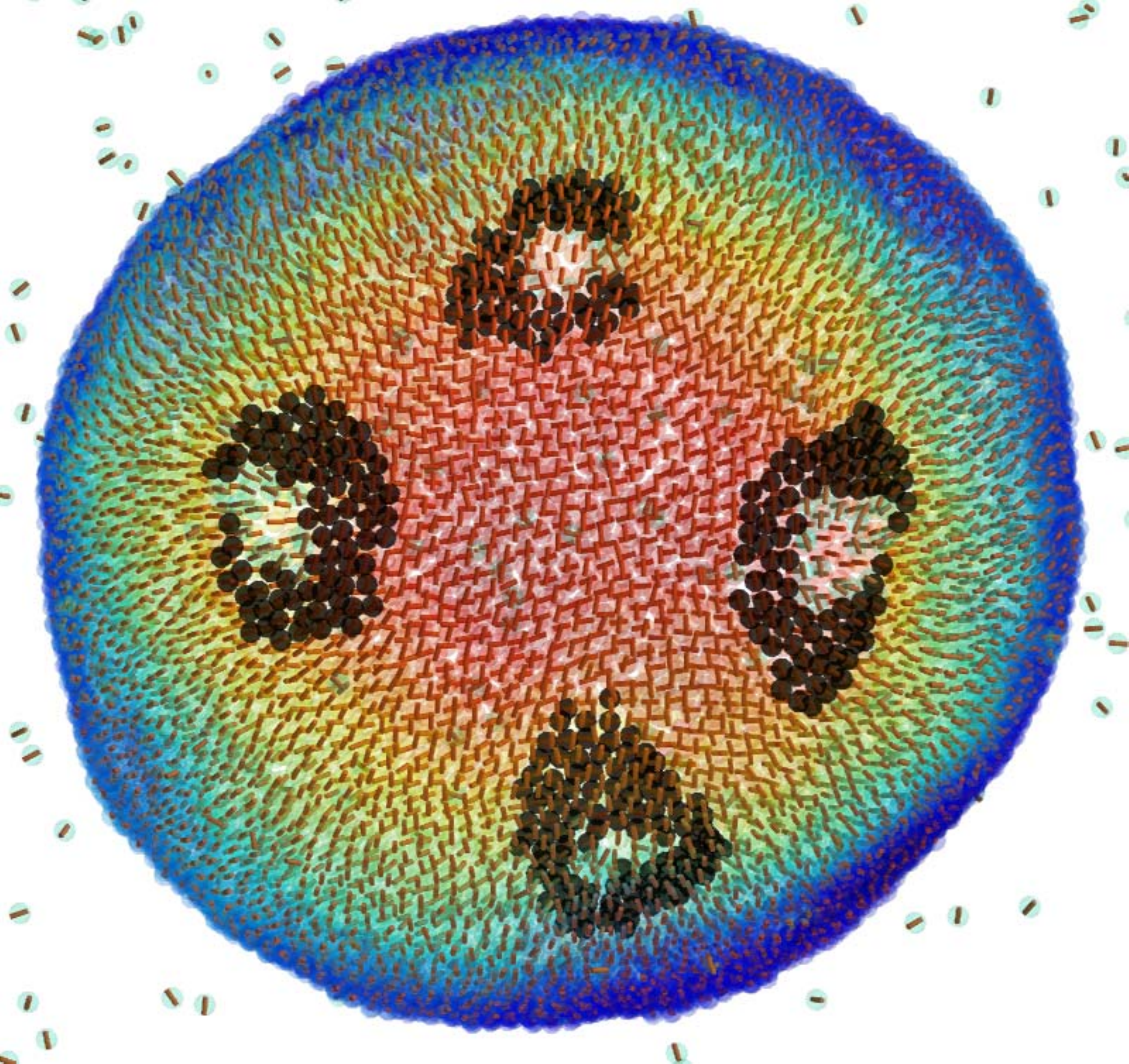} \\
  0.45 &
\includegraphics[width=0.96in]{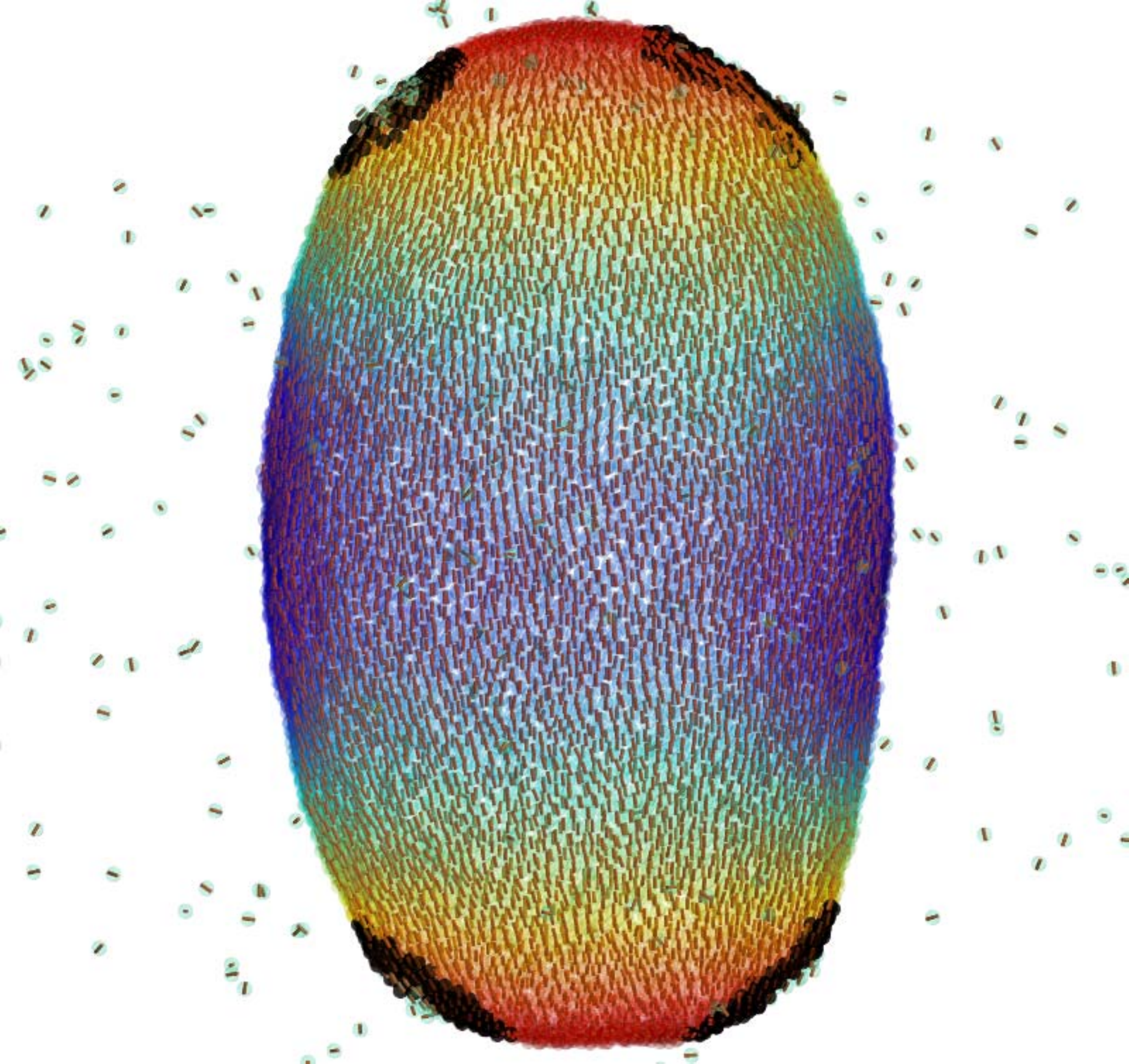} &
\includegraphics[width=0.96in]{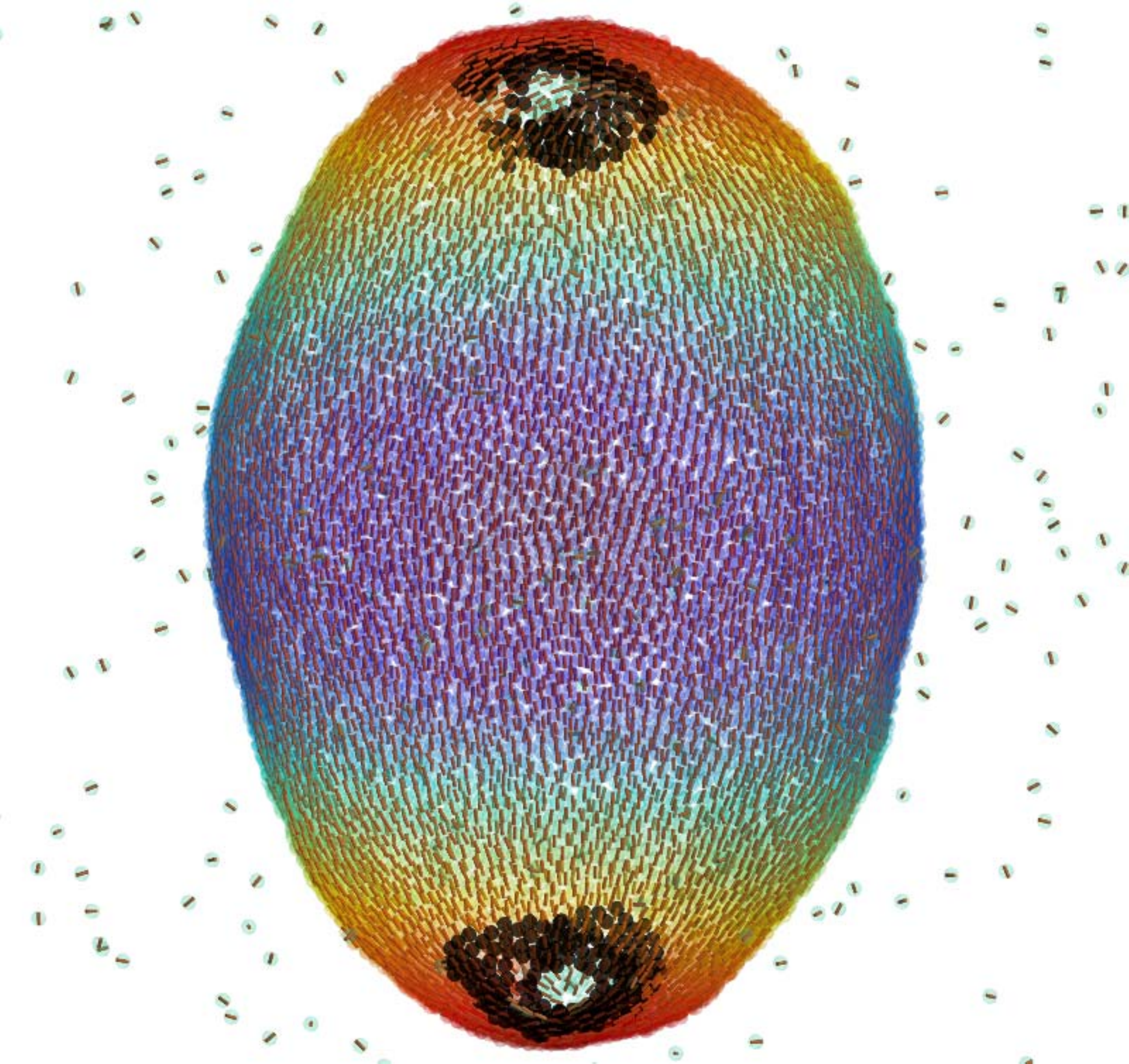} &
\includegraphics[width=0.96in]{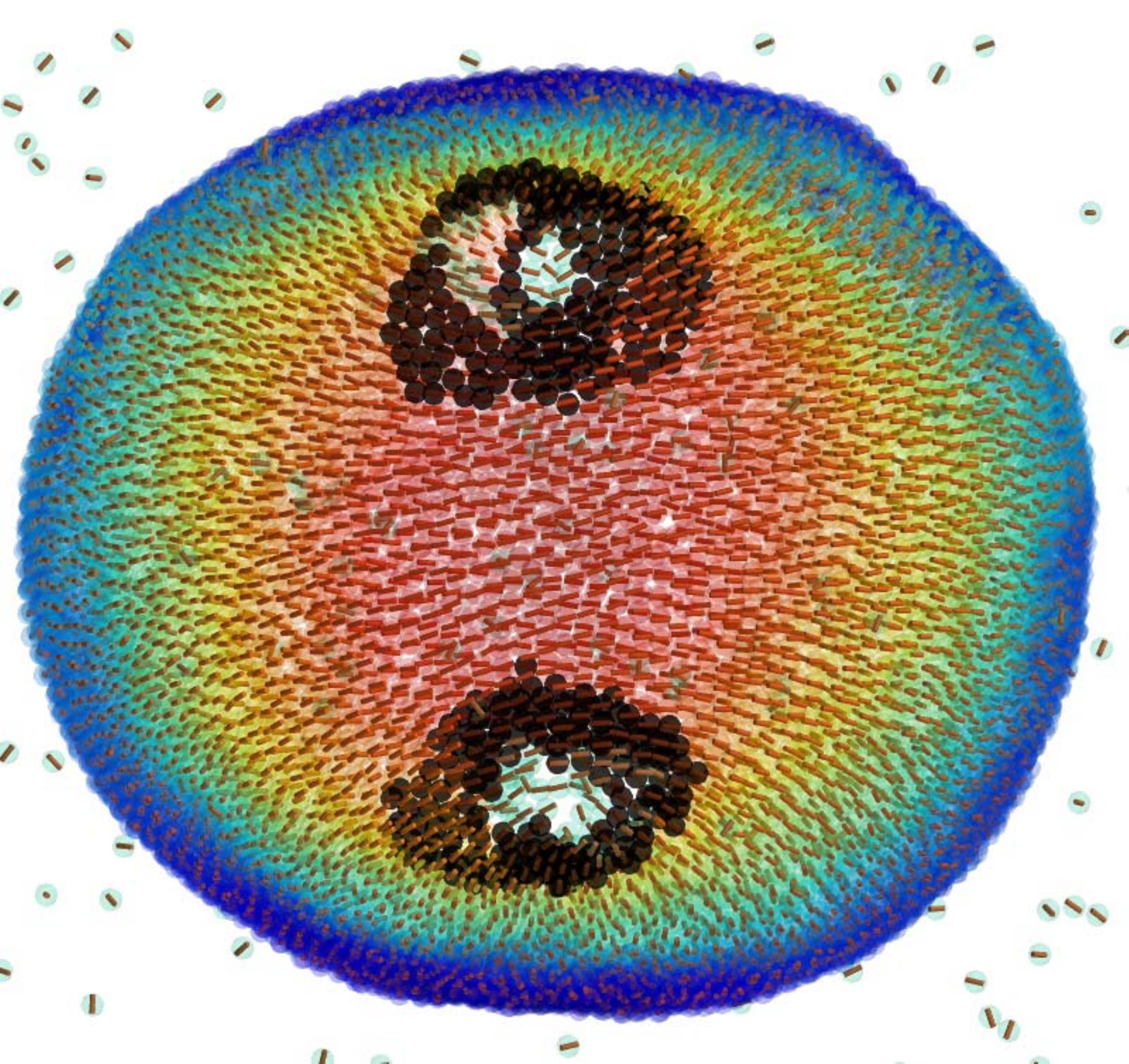} \\
  0.5 &
\includegraphics[width=0.96in]{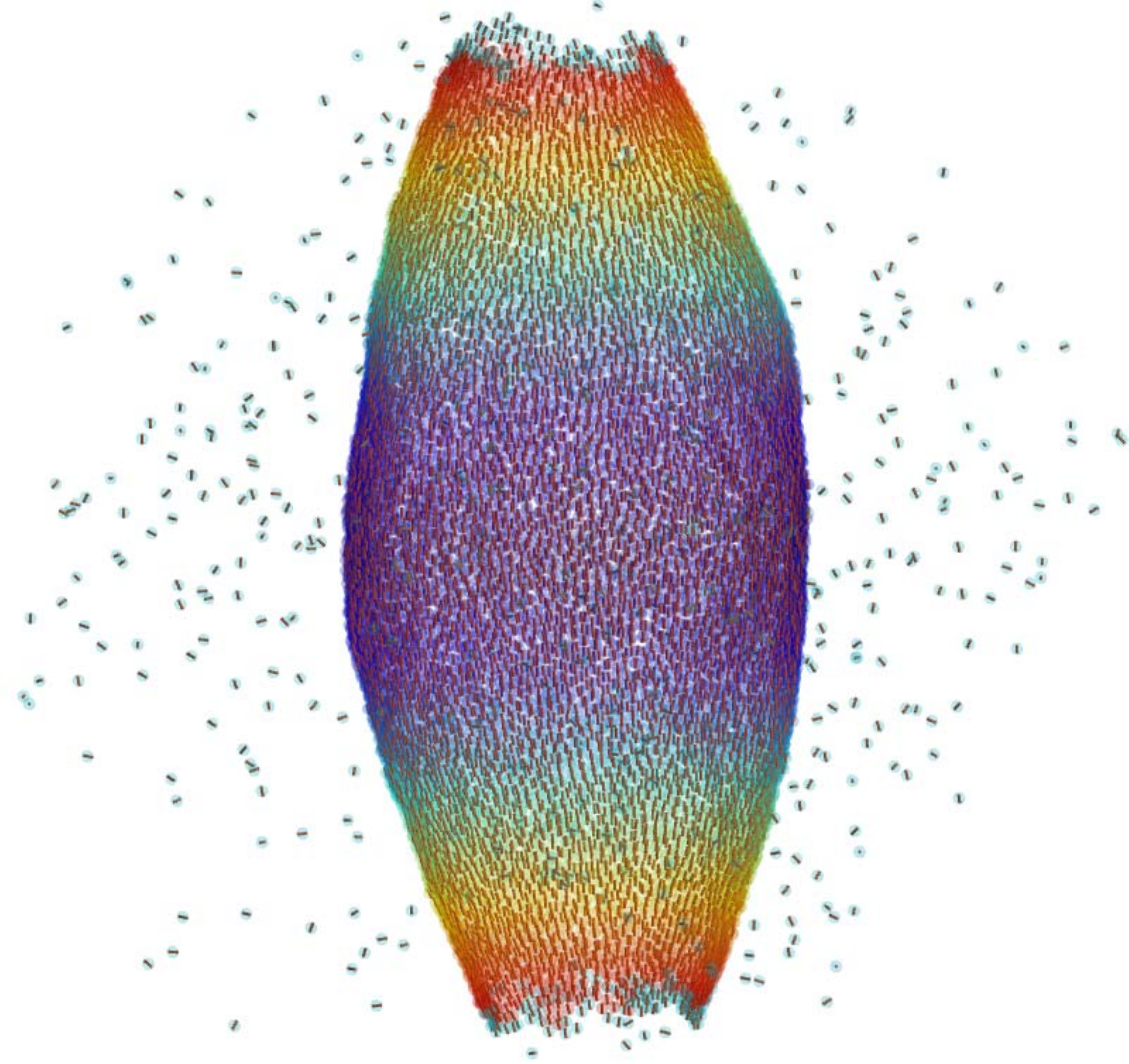} &
\includegraphics[width=0.96in]{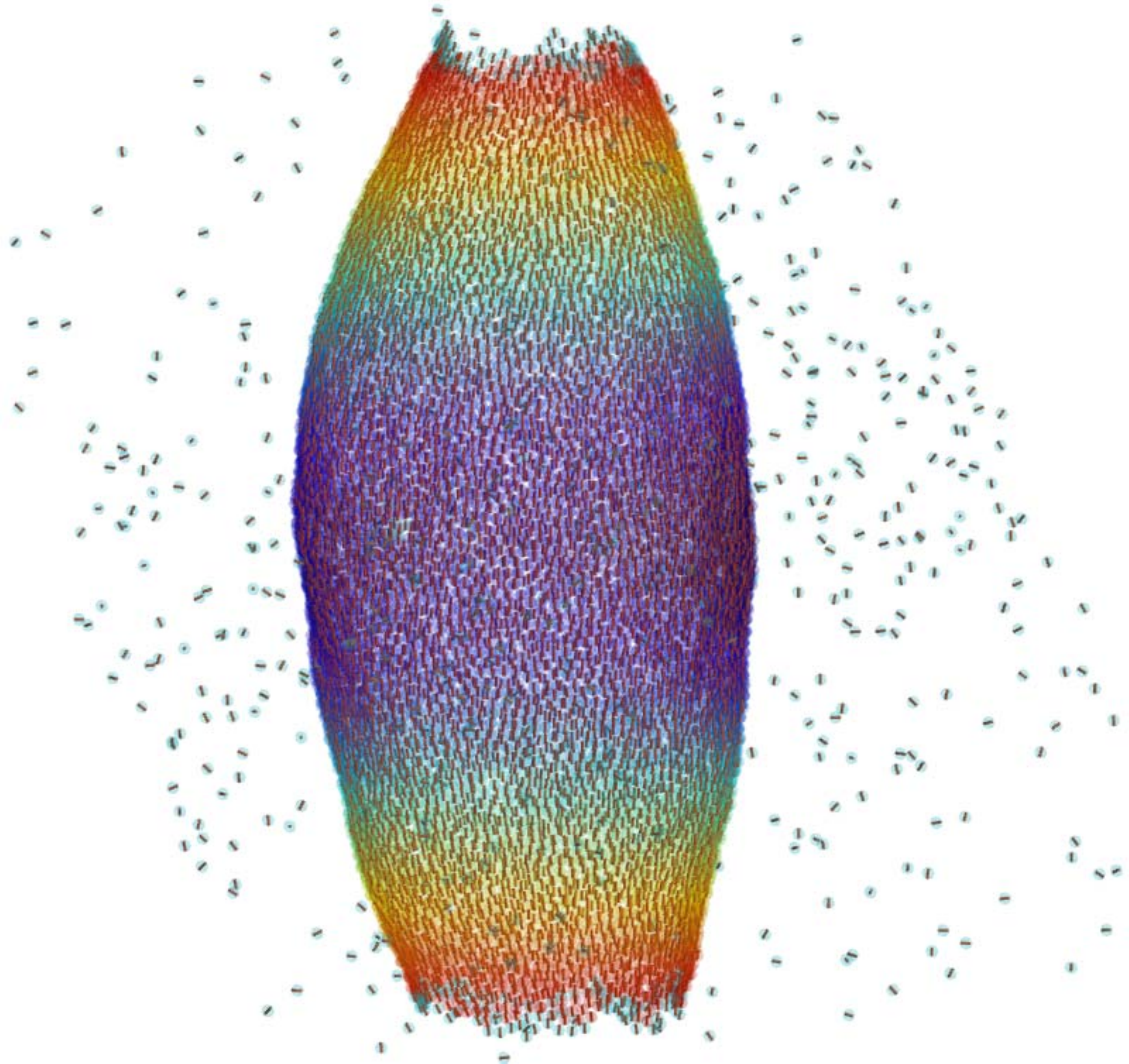} &
\includegraphics[width=0.96in]{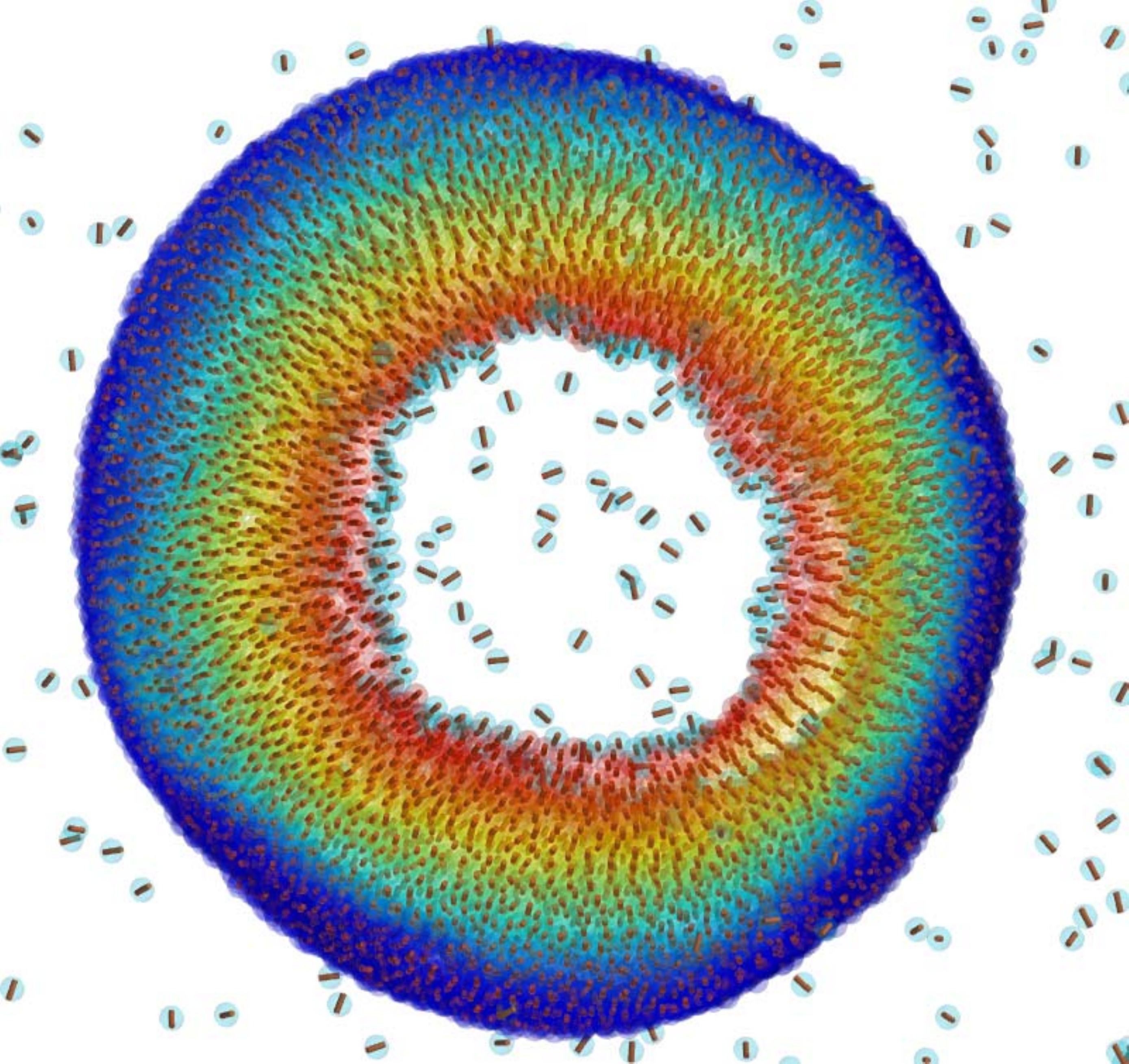} \\
\end{tabular}
\caption{\label{fig:shapes}(Color online) Simulated vesicles for several values of the parameter $\eta$. Color indicates distance from the vesicle's center of mass, and black dots indicate particles in or near a defect core. Particles are semi-transparent so that defects on both near/far sides are visible.}
\end{figure}

Varying the coupling parameter $\eta$ gives rise to a sequence of shape transitions. Snapshots of the front, side, and top views of the simulation results are shown in Fig.~\ref{fig:shapes}. Black dots indicate the locations of the four positive half-charged defects in the tangent plane (except for the largest $\eta=0.5$), and color indicates distance from the vesicle's center of mass. On the surface of the vesicles, the $\mathbf{\hat{c}}$ vectors are shown as line segments in the local tangent plane. The $\mathbf{\hat{n}}$ vectors are not shown. The coarse-grained particles are semi-transparent so that defects are visible on both the near and far sides of the vesicles.

From the snapshots in Fig.~\ref{fig:shapes}, we can see that changing $\eta$ changes the nematic director field and defect configuration as well as vesicle shape.   For the smallest $\eta=0.2$, the four defects align themselves approximately on a great circle. This great circle is consistent with the continuum prediction of Ref.~\cite{Shin2008} for nematic vesicles with a strong elastic anisotropy, so it may indicate that this potential gives a substantial difference between the effective Frank constants. As $\eta$ increases to 0.25, the defects move to form approximately a regular tetrahedron. For these small values of $\eta$, the vesicle shape is fairly close to spherical, indicating that the coupling between director distortions and curvature is not yet enough to substantially distort the sphere. As $\eta$ increases further to 0.30, the defects shift further, with two defects moving to each end of the vesicle. At the same time, the vesicle elongates substantially between these two poles, forming a prolate shape. For $\eta\ge 0.35$, the nematic order becomes stronger, so that the defect cores cost more energy. We then observe holes nucleating at the centers of the defects, and the vesicle coexists with a particle gas. The defect configuration also changes from an elongated tetrahedron ($\eta=$0.3--0.4) to rectangular ($\eta=0.45$). Finally, at the largest value $\eta=0.5$, each pair of defects fuses to form a large pore at each end of the vesicle, thus eliminating the defects and transforming the vesicle into a tube. The middle of the tube is still swollen, not a perfect cylinder, because of the model membrane's spontaneous curvature.

\begin{figure}
Front\includegraphics[width=3.2in]{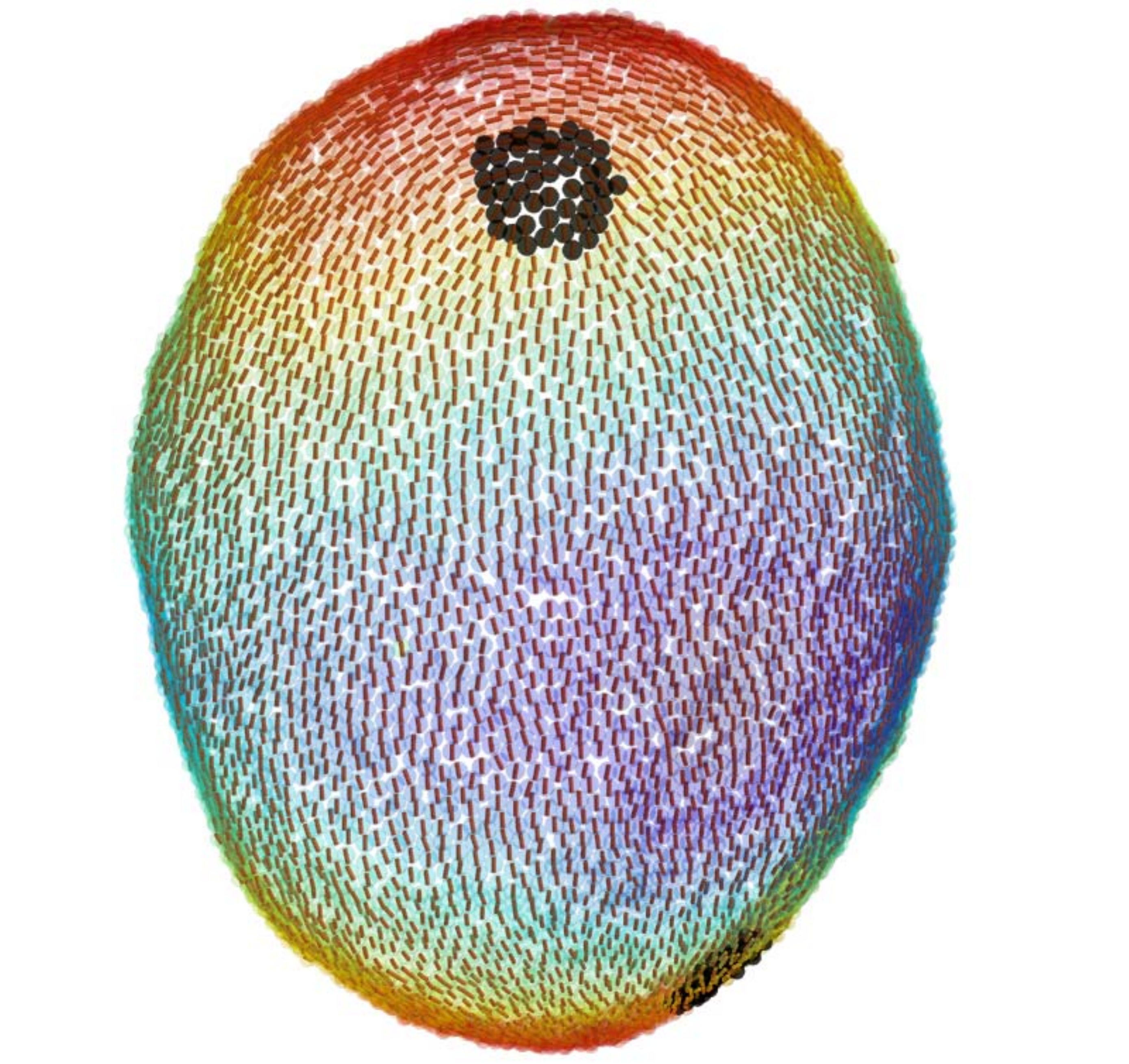}
Top\includegraphics[width=2.7in]{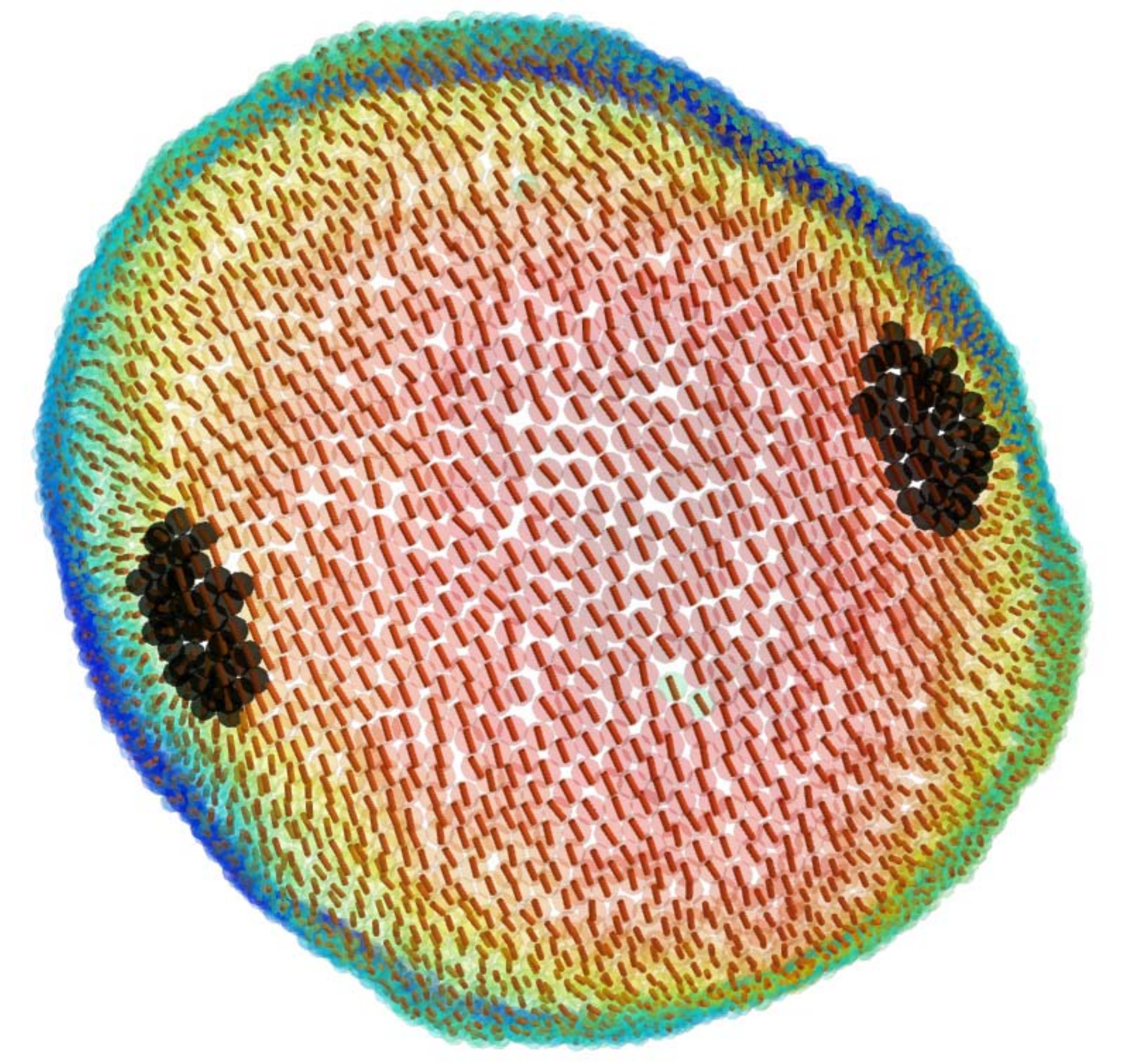}
\caption{\label{fig:zoom}(Color online) Enlarged front and top views of the vesicle for $\eta=0.3$, with line segments showing the local nematic orientation $\mathbf{\hat{c}}$ in relation to the overall shape. Particles are opaque; only defects on the near side are visible.}
\end{figure}

To show the relationship between the local nematic orientational $\mathbf{\hat{c}}$ and the overall shape more clearly, Fig.~\ref{fig:zoom} shows front and top views of the vesicle with $\eta=0.3$. In these views the coarse-grained particles are opaque, so that only two defects  (on the near side) are visible, and the defects can easily be identified as topological charge 1/2. From the front view, we see that the local nematic orientation $\mathbf{\hat{c}}$ is aligned along the long axis of the prolate vesicle, allowing the $\mathbf{\hat{c}}$ vectors of neighboring particles to be nearly  parallel in three dimensions (3D); transverse alignment would have much higher 3D interaction energy. From the top view, we see that the local nematic orientation $\mathbf{\hat{c}}$ is aligned perpendicular to the separation between the two defects. Hence, the director distortion between the defects is almost entirely bend rather than splay. Furthermore, the top view of the vesicle is not circular but extended along the average nematic director, perpendicular to the separation between the defects. Thus, the overall vesicle has a biaxial, potato-like shape.

It is remarkable that the half-charged defects are arranged in pairs, with a pair at each end of the vesicle, in spite of the usual repulsion between defects. We speculate that this arrangement occurs because the defects are attracted to the regions of high positive Gaussian curvature at each end of the vesicle. This attraction to the curved regions competes with the repulsion between defects to favor an optimum separation between the defects, which depends on the coupling parameter $\eta$. A similar pairing of defects has been seen in analytic calculations by Kralj~\cite{Kralj2010} for the optimum positions of nematic defects on colloidal particles with a fixed ellipsoidal shape. Here we see that the pairing occurs even when the shape is not fixed but is free to deform.

Our simulations can be compared with the predictions of Park \emph{et al.}~\cite{Park1992}, who performed analytic calculations of the shapes of deformable membranes with nematic and general $n$-atic order (note $n=2$ for a nematic phase). Their theory predicts that a nematic vesicle should deform into a shape with the symmetry of a regular tetrahedron, with a half-charged defect at each vertex. By contrast, our simulations never show the tetrahedral deformation, but only the extended biaxial, potato-like shape. We speculate that this discrepancy occurs because their theory is idealized in two ways. First, their free-energy functional considers only intrinsic coupling of director variations with curvature through a covariant derivative, explicitly neglecting other couplings allowed by symmetry~\cite{Peliti1989}. The interaction potential in our simulation includes an extrinsic coupling of the nematic director to the 3D curvature direction. In previous work we showed that the extrinsic coupling greatly changes the director field on surfaces with fixed curvature~\cite{Selinger2011}; here we see that it also affects shapes of deformable membranes. This effect may explain why the vesicle becomes extended along its long axis and why its ends become extended in a biaxial way; both distortions reduce the interaction energy of $\mathbf{\hat{c}}$ vectors in 3D. Second, their free-energy functional makes the approximation of a single Frank elastic constant, while our interparticle potential presumably gives different effective Frank constants. Shin \emph{et al.}~\cite{Shin2008} showed that anisotropy of Frank constants changes the arrangement of defects; here it also affects the membrane shape.

The nucleation of pores in a bilayer membrane depends on processes at length scales well below the particle spacing in our coarse-grained model~\cite{Ting2011}. While our coarse-grained simulation results demonstrate qualitatively that pores spontaneously nucleate at topological defects, more detailed theoretical analysis and molecular-scale simulations are needed to understand the process in detail. Another question is whether total enclosed volume is conserved during vesicle shape relaxation. The formation of pores in our simulation suggests that fluid may leak into or out of the vesicle as it evolves in shape, so we have not imposed a condition of fixed volume.

In conclusion, we have developed a coarse-grained particle-based model for simulating membranes with orientational order, and we have used it to study vesicles in the nematic liquid-crystal phase. The simulation results show surprisingly complex vesicle shapes and defect configurations, which arise from features in the interparticle potential. Thus, the simulation method enables us to explore the range of phenomena that can occur in soft materials where geometry interacts with orientational order and topological defects.

We thank A. Travesset for helpful discussions. Work was supported by NSF DMR-0605889 and 1106014.

\bibliography{CGNematicShell}
\end{document}